\RequirePackage{fix-cm}
\documentclass[smallextended]{svjour3}       
\smartqed  
\usepackage{hyperref}
\usepackage{natbib}
\usepackage{url}
\makeatletter
\def\url@leostyle{%
    \def\UrlFont{\sf}}{\def\UrlFont{\small\ttfamily}}
\makeatother
\usepackage{amsmath}
\usepackage{amsfonts}

\newcommand{\ZT}[1]{\textquotedblleft#1\textquotedblright}%
\newcommand{\define}{=_{\mathrm{def}}}
\usepackage{graphicx}
\usepackage{seqsplit}
\usepackage[normalem]{ulem}
\usepackage{color}
\begin{document}

\title{The Mereology of Thermodynamic Equilibrium
}


\author{Michael te Vrugt}


\institute{Michael te Vrugt \at
Institut f\"ur Theoretische Physik\\
Center for Soft Nanoscience\\
Philosophisches Seminar\\
Westf\"alische Wilhelms-Universit\"at M\"unster\\
D-48149 M\"unster, Germany
\email{michael.tevrugt@uni-muenster.de}           
}

\date{}

\maketitle

\begin{abstract}
The special composition question (SCQ), which asks under which conditions objects compose a further object, establishes a central debate in modern metaphysics. Recent successes of inductive metaphysics, which studies the implications of the natural sciences for metaphysical problems, suggest that insights into the SCQ can be gained by investigating the physics of composite systems. In this work, I show that the minus first law of thermodynamics, which is concerned with the approach to equilibrium, leads to a new approach to the SCQ, the \textit{thermodynamic composition principle} (TCP): Multiple systems in (generalized) thermal contact compose a single system. This principle, which is justified based on a systematic classification of possible mereological models for thermodynamic systems, might form the basis of an inductive argument for universalism. A formal analysis of the TCP is provided on the basis of mereotopology, which is a combination of mereology and topology. Here, \ZT{thermal contact} can be analyzed using the mereotopological predicate \ZT{self-connectedness}. Self-connectedness has to be defined in terms of mereological sums to ensure that scattered objects cannot be self-connected.

\keywords{Thermodynamics \and Mereology \and Special Composition Question \and Inductive Metaphysics \and Mereotopology}
\end{abstract}
\section{\label{intro}Introduction}
The \textit{special composition question} (SCQ) \citep{vanInwagen1987}, which asks under which conditions objects compose a further object, is among the central problems of modern metaphysics. It arises in \textit{mereology}, which is the theory of parthood relations \citep{BurkhardtEtAl2017}. In recent years, various attempts have been made to use results from the natural sciences to get insights into the SCQ. For example, \citet{McKenzieM2017} and \citet{HusmannN2018} have proposed physical bonding as a condition for composition, and \citet{CalosiT2014} have investigated the implications of quantum mechanics for mereology. Such approaches belong to the tradition of \ZT{inductive metaphysics}\footnote{It is difficult to distinguish between \ZT{inductive metaphysics} and the related concept of \ZT{naturalistic} or \ZT{scientific metaphysics}. Inductive metaphysics (see \citet{Scholz2019}) is an older and moderate tradition, associated with authors such as Gustav Theodor Fechner and Erich Becher. Naturalistic metaphysics \citep{LadymanR2007} is younger and essentially aims at reducing metaphysics to science. A discussion of the difference can be found in \citet{EngelhardFGS2021}. For this work, the difference is not central, I will be using the name \ZT{inductive metaphysics} throughout the article for consistency.} \citep{Scholz2019}, which aims to solve metaphysical problems based on empirical results \citep{Seide2020}. Inductive metaphysics has become a highly active field of research in philosophy that has provided interesting insights into problems such as the nature of time \citep{Saunders2002} or social metaphysics \citep{Scholz2018}. This makes scientific approaches to the SCQ a very promising route of inquiry.

There is, however, a fundamental theory of nature whose potential for the SCQ has not been fully explored yet, namely \textit{thermodynamics}. This is a branch of physics which studies certain properties of macroscopic objects, such as temperature, energy or entropy. It is among the central fields of physics and has thus also attracted the attention of metaphysicians, including those interested in mereology \citep{Needham2010}. Many of the systems that are studied in thermodynamics are composed of multiple macroscopic objects. It has been argued that it is among the general principles of thermodynamics to view multiple systems as a single system under the condition of thermal contact \citep[p. 286]{Wallace2015}. Such a principle, while being routinely employed in thermal physics, is not metaphysically innocent: If it holds, it provides at least a partial answer to the SCQ, such that studying the thermodynamics of composite systems is a very promising route for philosophers interested in inductive metaphysics.

The aforementioned principle is deeply connected to the concept of \textit{thermodynamic equilibrium} (see section (\ref{thermo})), which is introduced in thermodynamics through the minus first \citep{BrownU2001} and the zeroth law. As I will show, the idea of thermodynamic equilibrium is the main reason why thermodynamics is relevant for the SCQ. A good conceptual understanding of thermodynamic equilibrium is required not only for the metaphysical problem at hand, but also for other debates in philosophy of physics, most notably for the explanation of the irreversible approach to equilibrium. As shown by \citet{teVrugt2020}, a lack of terminological clarity regarding thermodynamic equilibrium tends to make it much more difficult to explain why systems tend to reach an equilibrium state. Consequently, a careful discussion of the concept of thermodynamic equilibrium, as will be provided here, is not only beneficial for metaphysics, but also for philosophy of physics and for physics itself. 

In this work, I will show that thermodynamics gives a very strong argument for viewing thermal contact as a condition for physical composition. This condition will be referred to as the \ZT{thermodynamic composition principle} (TCP). Since thermal contact is a rather general condition, this might form the basis of an inductive argument for universalism. A logical analysis of the TCP requires a way to formalize the idea of \ZT{being in contact}, which is possible using \textit{mereotopology} (a combination of mereology and topology \citep{Smith1996}). It will be shown that the TCP is naturally incorporated into closed extensional mereotopology combined with the assumption that contact implies underlapping. The relation \ZT{thermal contact} can then be analyzed based on the self-connectedness predicate, where the latter should be defined based on mereological sums.

This article is structured as follows: In section (\ref{thermo}), the TCP will be introduced and discussed from a metaphysical point of view. The implications of thermodynamics for the logical formalism of mereotopology are discussed in section (\ref{compo}). I conclude in section (\ref{conclusion}).

\section{\label{thermo}Composition in thermodynamics - a metaphysical perspective}

\subsection{The special composition question}
Suppose you find a book on your table, and you ask yourself \ZT{How many objects are on the table?} Interestingly, a large variety of answers is available. One of course is \ZT{one}, since there is one book. However, you could also argue that the book has 100 pages, such that there are 101 objects (the book and all its pages, which you count separately). You could even say \ZT{there are millions of atoms that are arranged like a book}, without thereby committing to the existence of the book as an object in its own right. 

We have thus, from a rather innocent question, arrived at the difficult philosophical problem of parthood and composition: Do composite objects such as the book exist, or are they just arrangements of their parts? And if they exist, under which conditions? This problem is of significant interest in metaphysics: If \textit{any} two objects $x$ and $y$ compose an object $z$, this might force our ontology to include some relatively strange objects. Consider the case that $x$ is the Eiffel tower and $y$ is your underwear\footnote{This example is from \citet[p. 191]{CarrollM2010}.}. It is intuitively plausible to say that an object $z$ that is composed of the Eiffel tower and your underwear does not exist. In contrast, if we consider a collection of iron atoms bound to each other, the existence of an object \ZT{block of iron} composed of the iron atoms is plausible. Thus, we need to find out under which circumstances it is the case that $x$ and $y$ compose an object. This is the content of the so-called \ZT{special composition question} \citep{vanInwagen1987}: If we have multiple objects (commonly referred to as \ZT{the $x$s}), what are the conditions for the $x$s to compose something?

Various answers are possible (see \citet[pp. 184- 226]{CarrollM2010}), which can be classified into four categories:
\begin{enumerate}
\item Universalism: It is always the case that the $x$s compose something. From a formal point of view, this is a clean and elegant solution, which does, however, have some drastic ontological consequences (for example, there \textit{does} exist an object composed of the Eiffel tower and your underwear).
\item Nihilism: It is never the case that the $x$s compose something, except if there is only one of them. The only existing objects are mereological simples, i.e., objects that do not have proper parts. Again, this is a clean and elegant solution, and again, it has drastic consequences (for example, neither the Eiffel tower nor your underwear would exist, only Eiffel-tower-shaped and underwear-shaped arrangements of elementary particles).
\item Brutal composition: There is no general answer to the special composition question, the fact that the $x$s do or do not compose something is simply a brute fact.
\item Moderate compositionalism: There is a condition for the $x$s to compose something. In this case, we obviously need to specify what this condition is. Suggestions from the literature include constituting a life \citep{vanInwagen1995} or physical bonding \citep{McKenzieM2017,HusmannN2018}. 
\end{enumerate}

In this work, I aim to develop a partial answer to the SCQ based on results from thermodynamics. This is a promising route of inquiry: Various authors have approached the problem of composition from a scientific perspective \citep{Schaffer2010,Healey2013,CalosiT2014,McKenzieM2017,HusmannN2018}, and mereology has become an important topic in philosophical approaches to thermodynamics \citep{Needham2013,Needham2010b} as well as related fields such as chemistry \citep{Needham2007,HarreL2013} and biology \citep{JansenS2011,JansenS2014}. In particular, I will argue that the $x$s compose an object if they are in thermal contact. This is based on the fact that the combined system has an unique equilibrium state, which is a novel property compared to the (equilibrium states of) its constituents. My proposal is a form of moderate compositionalism, which, however, implies universalism in a universe in which all objects are in thermal contact.

\subsection{\label{laws}The laws of thermodynamics}

Thermodynamics is a phenomenological theory that describes macroscopic systems without explicitly considering their microscopic constituents (in contrast to statistical mechanics, which describes a macroscopic system based on the dynamics of the microscopic particles it consists of). Its principles have been confirmed by a huge variety of experiments, such that one can safely argue that they are sufficiently well established to form a basis for metaphysical considerations. Thermodynamics is based on five laws (\ZT{Haupts\"atze}), which are assumed to hold universally. These are the minus first law (spontaneous approach to equilibrium), the zeroth law (transitivity of equilibrium), the first law (conservation of energy), the second law (entropy cannot decrease) and the third law (zero temperature cannot be reached). The minus first law is not among the set of laws usually presented in physics textbooks, but is widely accepted in the philosophical foundations of physics as being necessary for completing the axiomatic system.

A central concept in thermodynamics is the idea of \ZT{thermodynamic equilibrium}. It has received significant attention in the philosophy of physics (see \citet{teVrugt2020} for a discussion), but is, as I will show here, also important for metaphysics. Of particular interest for the present discussion are the minus first and the zeroth law, which I will present here in more detail. The minus first law, introduced by \citet{BrownU2001}, is concerned with the approach to equilibrium: If a cup of hot coffee stands in a cold room, it will spontaneously cool down until it reaches room temperature, but it will never spontaneously heat up. This seems to be a universal fact in thermodynamics. Often, it is associated with the second law of thermodynamics, which states, very roughly, that a quantity known as \textit{entropy} cannot decrease in a closed system (see \citet{Uffink2001} for a more sophisticated discussion). However, according to \citet{BrownU2001}, the tendency of systems to equilibrate is not entirely captured by the second law. The minus first law states that for each isolated system there exists a unique\footnote{The equilibrium state is unique given certain external constraints. In the original statement \citep[p. 528]{BrownU2001}, one constraint is specified to be a finite fixed volume.} state of equilibrium that the system will spontaneously enter. This law introduces the idea of equilibrium on a macroscopic level. It then gives the observation that systems tend to approach an equilibrium state spontaneously the status of an axiom. What is interesting here is that \ZT{being in equilibrium} is a property of a single system. From the perspective of predicate logic, there seems to exist a one-place predicate \ZT{being in equilibrium}, which I will call $E_1$. Then, according to the minus first law, for any isolated system $x$, it will after a certain time be true that $E_1 x$. 

The zeroth law, in contrast, is not concerned with the approach to equilibrium but with the properties of the equilibrium state. It states that if a system $x$ is in equilibrium with a system $y$, and if system $y$ is in equilibrium with another system $z$, then the systems $x$ and $z$ will also be in equilibrium with each other. Equivalently, one can say that the zeroth law states the transitivity of thermodynamic equilibrium. Physically, two systems are in (thermal) equilibrium with each other if they do not exchange\footnote{Throughout this article, I will use the word \ZT{exchange} synonymously with \ZT{net exchange}. This means that the statement \ZT{systems $a$ and $b$ do not exchange energy} does not exclude that there are energy fluctuations, it simply means that the average energy of both systems is constant.} heat\footnote{For non-physicists, a brief discussion of the difference between \ZT{heat} and \ZT{temperature} is in order: The temperature, which can be connected to the mean kinetic energy of the particles the system consists of, depends on the state of the system. Heat is a form of energy that is exchanged between two systems with different temperatures that are in thermal contact.} despite being in thermal contact. The relation \ZT{being in thermal contact with} holds between two systems if they can exchange heat. From a logical point of view, it is notable that, while the minus first law talks about isolated systems that are in equilibrium, the zeroth law talks about systems in thermal contact that are in equilibrium \textit{with each other}. Thus, there seems to be a two-place-predicate (relation) \ZT{being in equilibrium with}, that I will call $E_2$. Logically, the zeroth law demands that
\begin{equation}
E_2 xy \land E_2 yz \rightarrow E_2 xz.
\label{zerothlaw}
\end{equation}
If $E_2 xy \land E_2 yz$ holds, we can assign the same temperature to the systems $x$,$y$ and $z$ (which is then a property that the individual systems possess).

It is important to distinguish between $E_1$ and $E_2$ (\ZT{being in equilibrium} and \ZT{being in equilibrium with}), both for physical and for logical reasons (a one-place predicate is obviously not the same as a two-place predicate). This distinction is, unfortunately, not always made with sufficient care in discussions of thermodynamics. A system satisfies $E_1$ if it is in the unique stationary state that it will spontaneously enter. Two systems satisfy $E_2$ if they are in thermal contact, but do not exchange heat. (Contrary to a widespread belief, the relation $E_2$ requires \textit{actual} thermal contact, i.e., it cannot be introduced as \ZT{would not exchange heat if brought into thermal contact}\footnote{As shown by \citet{HilbertHD2014}, this is problematic since the thermodynamic state of an isolated system is not always uniquely determined by the temperature. In general, knowing the microcanonical temperature of two systems does \textit{not} allow to predict heat flows between them.} \citep{HilbertHD2014}.)

Having discussed the general idea of thermodynamic equilibrium, we can now relate it to the metaphysical problem of composition. For this purpose, let us consider a discussion of the zeroth law by \citet[p. 54]{Wallace2018}:

\textit{\ZT{Multiple systems in (perhaps-generalised) thermal contact may be treated as a single system; in particular, any such combined system will have a unique equilibrium state. [...] The Second Law of Thermodynamics generalises to require that the total entropy of two systems in (perhaps-generalised) thermal contact does not decrease when those systems exchange energy and other conserved quantities. [...]  If two systems are in thermal contact, and heat $\delta Q$ flows from system 1 to system 2, the total change in entropy is $\delta S = \delta Q (1/T_2 - 1/T_1)$. So heat will flow only if $T_1 > T_2$, and indeed, no process can as its sole effect induce heat flow unless this condition holds (the Clausius statement of the Second Law). It follows that a necessary and sufficient condition for two systems in thermal contact to be jointly at equilibrium is that they are separately at equilibrium with equal temperatures. (This generalises to other forms of contact.) As a consequence, the relation ‘at equilibrium with’ is an equivalence relation: this is the Zeroth Law of thermodynamics, and in textbook presentations is often taken as a starting point; in my presentation, it is a consequence of other assumptions.}.}

In summary: Wallace introduces here the assumption that multiple systems, if brought in thermal contact, form a single system. This system will then, by the minus first law, have a unique equilibrium state which it spontaneously approaches. During this process, the total entropy cannot decrease (by the second law). This then implies that two systems in thermal contact will reach a state in which they are at equilibrium with each other and thus have equal temperatures. 

What still needs to be clarified is the notion of \ZT{generalized thermal contact}. As explained above, \ZT{thermal contact} denotes the ability to exchange heat. If two systems are in thermal contact, but do not exchange energy, they are said to be in thermal equilibrium with each other. Thermodynamic systems can also exchange other conserved quantities, such as volume. This can be incorporated using the idea of \ZT{generalized thermal contact}. For example, if two systems do not exchange volume despite having the ability to do so, they are in pressure equilibrium \citep{HilbertHD2014}. The argument is based on the minus first law, which predicts not only that systems equalize temperatures if they are in thermal contact, but also that they equalize pressures if they are able to exchange volume.\footnote{Strictly speaking, the minus first law only predicts this once we assume that the equilibrium state it introduces is characterized by, e.g., equal temperatures of systems in thermal contact. This, however, is an assumption that is known to be true from other axioms and from experiment.} Thus, our analysis should be based on generalized thermal contact. For most of this work, I will nevertheless speak of \ZT{thermal contact} for brevity.

While technical details of Wallace's discussion cited above are not of central importance here, the first assumption is (despite the innocent formularion \ZT{may be treated}) very interesting and far from being metaphysically innocent: \ZT{Multiple systems in (perhaps-generalized) thermal contact may be treated as a single system} can - although intended here to be a purely physical statement - be seen as a partial answer to the SCQ, as it states the existence of composite systems and provides a criterion for composition (namely thermal contact). And, very notably, this composition principle is not only important for the philosopher in the armchair\footnote{To any philosopher offended by this cliche, I wish to point out that there is nothing wrong with sitting in an armchair.}, it is of actual physical relevance, since it is required to derive further assumptions about the nature of thermodynamic equilibrium which are essential for modern thermodynamics.

\subsection{Mereological models of thermodynamics}

It is common practice in many scientific disciplines to discuss systems consisting of multiple objects. Thus, some discussion is required to assess whether the system composed of multiple systems in thermal contact does exist as a metaphysically relevant composite object. A common view is that a strong case for the existence of a composite system can be made if it has a property that cannot be reduced to properties of its constituents or relations between them. Views of this form have been expressed by \citet{CalosiT2014} and \citet{Maudlin1998} in their discussion of composite quantum systems, who argued that the property \ZT{being in an entangled state} of composite quantum objects cannot be reduced in such a way. If a system has such an irreducible property, we have a good argument for its existence as a composite object.

In our case, the property $E_1$ of the composite system appears to be a candidate for such a property. To see this, consider an isolated system of two boxes $x$ and $y$ in thermal contact. After a sufficiently long time, these boxes will have equal temperatures, i.e., they are in equilibrium \textit{with each other} ($E_2xy$). On Wallace's account, one would view these two boxes as a single system\footnote{Here, the + sign indicates a mereological sum (see section (\ref{compo})).} $z = x+y$. This joint system then satisfies $E_1 z$, i.e., it is in equilibrium. $E_1 z$ means, by the minus first law, that $z$ is in the unique state that $z$ spontaneously approaches. This unique state, and thus the property $E_1$, is a property of the joint system. 

It might now be argued that the statement $E_1 z$ about the joint system is here simply another way of saying that $E_2xy$, since the joint system is in equilibrium because its subsystems are in equilibrium with each other. In this case, the statement $E_1 z$ about the joint system could be fully reduced to statements about the subsystems, such that the system $z$ would not be required. A view of this form was (although not in a metaphysical context) expressed by \citet{HilbertHD2014}:

\textit{\ZT{Two systems are in thermal equilibrium, if and only if they are in contact so they can exchange energy, and they have relaxed to a state in which there is no average net transfer of energy between them anymore. A system $\mathcal{A}$ is in thermal equilibrium with itself, if and only if all its subsystems are in thermal equilibrium with each other. In this case, $\mathcal{A}$ is called a (thermal) equilibrium system.}} 

There is, however, an important reason for why the system $z$ \textit{is} needed: The minus first law predicts that any \textit{isolated} system $z$ will, after a certain time, satisfy $E_1 z$. In contrast, the laws of thermodynamics do not directly involve the statement that the relation $E_2$ will hold between two systems $x$ and $y$ after a certain time once they are placed in thermal contact (since the minus first law, which captures the tendency to approach thermodynamic equilibrium, is concerned with $E_1$ and not with $E_2$). The reason thermodynamics nevertheless predicts that two systems $x$ and $y$ will reach equal temperatures is that the minus first law holds for the system $z=x+y$, that it predicts that $E_1 z$ will hold after a certain time, and that it follows from $E_1 z$ that $E_2 xy$. If the joint system $z$ would not exist, thermodynamics would not predict $x$ and $y$ to be in thermal equilibrium with each other after a certain time. Since thermodynamics obviously does predict this, it requires the existence of the joint system. 

A systematic way to address this point in more detail is to first classify all possible mereological models for systems with irreducible properties, which then enables one to discuss all of them in order to identify the most plausible one. The first problem has been addressed for the case of entangled\footnote{A state of a quantum system is said to be entangled if it cannot be reduced to the states of its constituents (formally: if its wavefunction cannot be written as a product of the wavefunctions of the constituents).} quantum systems by \citet{NaegerS2020} and \citet{Naeger2020}. They have identified six possible mereological models, which can (with little modification\footnote{In \citet{NaegerS2020} and \citet{Naeger2020}, the property of interest is \ZT{being in an entangled quantum state}. This property can then be monistic, relational or pluralistic. Here, of course, the property we are concerned with is $E_1$.} ) also be applied in the case of thermodynamics:
\begin{enumerate}
    \item Moderate monistic holism (MMH): $E_1$ is a property of the combined system, the subsystems also exist.
    \item Radical monistic holism (RMH): $E_1$ is a property of the combined system, the subsystems do not exist.
    \item Moderate relational holism (MRH): $E_1$ can be reduced to a relation $E_2$ holding between the subsystems, the combined systems also exists.
    \item Radical relational holism (RRH): $E_1$ can be reduced to a relation $E_2$ holding between the subsystems, the combined system does not exist.
    \item Moderate pluralistic holism (MPH): The property $E_1$ is carried collectively by the subsystems, the combined system also exists.
    \item Radical pluralistic holism (RPH): The property $E_1$ is carried collectively by the subsystems, the combined system does not exist.
\end{enumerate}
In the following, I will discuss all of these approaches and thereby show that MMH is the most plausible one in the context of thermodynamics.

First, we can rule out the options MRH and RRH. Since $E_2$ is the most plausible candidate for a relation that $E_1$ can be reduced to, and the only relation that is usually suggested to serve this function, we do this by discussing why this particular reduction is not possible (although it is logically possible that there is another relation that does allow for such a reduction). For this purpose, we consider the case of \textit{active matter}. Active systems consist of particles that convert energy into directed motion, such as swimming bacteria. For these systems, it is characteristic that energy enters the system at the level of each particle rather than just through the boundary of the system \citep[pp. 1144-1145]{MarchettiJRLPRS2013}. Consequently, the parts of an active system do not necessarily exchange energy with each other \footnote{They do, however, exchange energy with the environment; active systems are not isolated. This is why a definition of $E_1$ based on the minus first law, which applies to isolated systems, correctly identifies stationary active systems as nonequilibrium systems.}. Let us now assume that we, as suggested above, define $E_2xy$ as \ZT{There is no net energy exchange between $x$ and $y$}. \textit{If} we use such a definition of the relation $E_2$, then it would follow that the two active systems are in equilibrium with each other.

Nevertheless, active systems are not in equilibrium. This is reflected by a variety of phenomena, such as \textit{motility-induced phase separation} \citep{CatesT2015}, where particles accumulate even if they only have repulsive interactions \citep{BickmannW2019b}. This is impossible in equilibrium systems.  An even better example in the present context is \textit{motility-induced temperature difference}: As shown by \citet{MandalLL2019}, it is possible in active matter that phases with different (kinetic) temperatures are in coexistence\footnote{This effect requires the active particles to be underdamped \citep{MandalLL2019}. Underdamped active matter can exhibit a variety of effects not present in overdamped active systems (see \citet{Loewen2020,teVrugtJW2021} and references therein).}. 

It is not thermodynamically reasonable to say that two systems $x$ and $y$ are in equilibrium with each other if they are not separately in equilibrium. Formally, $E_2xy$ should imply $E_1 x \land E_1 y$. The example of motility-induced temperature difference shows this very clearly: In equilibrium thermodynamics, the relation $E_2xy$ is used to introduce the idea of a thermodynamic temperature (if $E_2xy$ holds, then $x$ and $y$ have the same temperature). Thus, we should not define $E_2$ in such a way that it can hold between two systems with different temperatures. If, however, $E_1x \land E_1 y$ is a part of the definition of $E_2xy$, then it would be circular to define $E_1$ in terms of $E_2$. Consequently, relational approaches to $E_1$ are not a viable option.

Next, we consider MPH and RPH, which have in common that $E_1$ is assumed to be a property that is carried collectively by the constituents of the equilibrium system (implying that we do not necessarily have to assume the existence of the combined system). \citet{Naeger2020} mentions \ZT{having temperature 12 $^\circ$C} as an example for such a property - if we assume temperature to be proportional to the mean kinetic energy of the particles, the property of having a certain temperature is carried collectively by the individual particles. He argues that \ZT{being in an entangled state} is a property that is carried collectively by a system of quantum-mechanical particles. In particular, N\"ager prefers RPH over MPH based on the argument that the macro objects are not required in the theory (they are not causally relevant and their time-independent properties straightforwardly reduce to those of their constituents). The idea of properties instantiated collectively has also been put forward in the context of quantum entanglement by \citet[p. 667]{Brenner2018} to establish the view that one cannot infer the metaphysical reality of composite objects from the fact that these objects are used in science.

However, including composite systems into our ontology can add explanatory value. \citet[p. 671]{Brenner2018} discusses such an argument for the case of entanglement: The fact that a composite quantum system is in a singlet\footnote{For non-physicists: In a system of two quantum-mechanical particles in a singlet state, the particles have opposite spins and are entangled.} state can \textit{explain} (in the sense of a grounding relation) why the individual particles are entangled, an idea put forward by \citet{IsmaelS2020}. This argument can also be applied here: As discussed above, the fact that the composite thermodynamic system is in the state $E_1$ can explain why the constituents stand in the relation $E_2$, and the minus first law (demanding that an isolated system always approaches equilibrium in the sense of $E_1$), applied to the composite system, can explain why systems in thermal contact will, after a certain time, stand in the relation $E_2$. Brenner rejects this proposal for the case of entanglement based on (a) the ontological cost of assuming composite systems and (b) the fact that entanglement relations are also possible between different degrees of freedom of the same object, where composition is not required as an explanation. Regarding (a), we can however object that not including entities in our ontology that improve our scientific explanations due to metaphysical concerns is a misapplication of the principle of parsimony. Moreover, the objection (b) is developed for the entanglement case and not applicable here: The explanation for why a non-composite system (supposing thermodynamics is relevant for such systems) approaches equilibrium is simply the same as for a composite system, namely that the minus first law demands it.
 
If $E_1$ is neither reducible to a relation nor carried collectively, the only options left are RMH and MMH. We thus have to analyze whether it is plausible to assume that \textit{only} the composite system exists (as stated by RMH). This, however, is not plausible since, in this case, thermodynamics of \ZT{composite} systems would be identical to thermodynamics for single systems. There are, however, important differences, such as the fact the fact that thermodynamics for composite systems places significantly more restrictions on the definition of the entropy \citep[p. 54]{Wallace2018}. Moreover, \textit{if} we have an equilibrium system satisfying $E_1$, it is certainly true that its subsystems satisfy $E_2$. This fact, like the minus first law, is essential to thermodynamics since it forms the basis of the zeroth law and therefore of the idea of a thermodynamic temperature. Consequently, the existence of the composite system is not sufficient, we also require the existence of the subsystems. This rules out RMH, leaving MMH as the only option.

This settles the issue if the composite system has only two parts. For systems with three or more parts, however, MMH can take various forms. To see this, consider an isolated system consisting of three boxes $a$, $b$ and $b$ in thermal contact. The boxes $a$ and $c$ are connected via the box $b$, such that if $b$ is removed, $a$ and $b$ are no longer in thermal contact. According to MMH, there exists a system consisting of all three boxes (that will approach equilibrium). What is not determined by the previous line of argument is whether there exists a system consisting of only two of the boxes, such as $a$ and $b$ or $a$ and $c$. Hence, there are two possible forms of MMH: MMH$_1$, including only $a$, $b$, $c$ and the total system, and MMH$_2$ including also any combination of subsystems.

To see why we require MMH$_2$, note that the unique equilibrium state of an isolated system is determined by its conserved quantities and external constraints \citep[p. 53]{Wallace2018}. Thus, the property \ZT{having a unique equilibrium state} does not depend on whether the system actually realizes this state, i.e., whether it actually satisfies $E_1$ (otherwise the composite system might not exist until it reaches equilibrium, which would be a strange consequence). What matters is that the system has such an unique equilibrium state, and that it is a property of the combined system. In our case, the system consisting only of boxes $a$ and $b$ also has such an equilibrium state (determined by the conserved quantities of these two systems), which it approaches when box $c$ is removed. This implies the existence of a system consisting of $a$ and $b$. We can even run this argument for the system consisting of $a$ and $c$: Suppose that there can be no heat exchange with box $b$, and that box $b$ has a fixed volume (in contrast to boxes $a$ and $c$). In this case, volume exchange between $a$ and $c$ is still possible, such that they are in generalized thermal contact. Thus, the system consisting of $a$ and $c$ will reach an equilibrium state (in which they have equal pressure). Since this state depends on the properties of $a$ and $c$ ($b$ merely establishes contact\footnote{A thermodynamic system is isolated if no heat or matter exchange with the environment is possible \citep[]{HilbertHD2014}. Consequently, the presence of $b$ does not imply that the system consisting of $a$ and $c$ is not isolated.}), it is a property of the system consisting of $a$ and $c$, which thus also has to exist. Note, however, that if we remove $b$ entirely, there is no longer any form of thermal contact between $a$ and $c$, such that they just separately reach their own equilibrium state. In this case, thermodynamics provides no argument for the existence of the joint system.

There is a further and simpler reason why the existence of subsystems is also required: While the minus first law requires isolated systems to approach an equilibrium state, it is also possible for non-isolated systems to be in equilibrium. In particular, a system $a$ might be in equilibrium ($E_1 a$) while being in contact with a system $b$ that is not. Since the equilibrium state is a property of the system $a$, the system $a$ should exist (even if it consists of multiple systems). \textit{Interventionists} \citep{RidderbosR1998}, who explain the approach to thermodynamic equilibrium by perturbations from the environment, even hold that only non-isolated systems can reach equilibrium.\footnote{This implies that the minus first law cannot be strictly true for an interventionist \citep{teVrugt2020}. However, the predicate $E_1$ can be defined in other ways based on the microscopic configuration of the system. Note that I am not attempting to defend interventionism here, since my argument is based on the minus first law.}

We are thus in the position to give a partial answer to the SCQ, which I will call the \ZT{thermodynamic composition principle} (TCP):
\begin{itemize}
\item $\exists y$ (the $x$s compose  $y$) if they are in (generalized) thermal contact.
\end{itemize}
Note that the TCP contains an \ZT{if} and not an \ZT{iff}, i.e., it does not rule out other physically motivated composition principles based, e.g., on quantum entanglement or bonding.

\subsection{Objections and implications}

Next, I discuss possible objections to and metaphysical implications of the TCP. Contact as a condition for composition is not very popular. In fact, many authors discussing the SCQ start by presenting contact as an example for a \ZT{wrong} answer \citep[pp. 26-29]{vanInwagen1987} before presenting what they consider to be the \ZT{right} answer. A first thing to note here is that the TCP is based on thermal contact, which is not the same as geometrical contact. For example, two objects that do not touch each other (no geometrical contact) may still exchange heat by thermal radiation (thermal contact). Nevertheless, the two types of contact are certainly linked (objects that touch each other will typically also exchange heat), such that some objections to the idea of contact as a composition criterion should also be discussed here.

It is helpful at this point to discuss these two types of contact. First, we have geometrical contact (often referred to as physical contact). Two objects are in geometrical contact if they touch each other. This is a very intuitive definition that, however, is rather difficult to formalize - on the level of elementary particles, two (macroscopic or microscopic) objects will not be in geometrical contact in any meaningful way \citep[p. 34]{vanInwagen1995}. A possible workaround for macroscopic objects would be that they are said to be in contact if the distance between their surfaces decreases below a certain value (say, a few atomic diameters). Second, there is generalized thermal contact, which was introduced in section (\ref{laws}). In the terminology used here, two objects are in generalized thermal contact if they can exchange conserved quantities. The most important such quantity is energy, and two objects are in thermal equilibrium if they do not exchange heat despite being able to do so. However, as mentioned above, other conserved quantities may also play a role here. One such quantity is particle number, which can be exchanged by diffusion. If systems do not exchange particles despite being able to do so, they are said to be in chemical equilibrium \citep{HilbertHD2014}. All of the above considerations also apply here. In particular, it is still possible to view two macroscopic objects that exchange particles as distinct (i.e., to accept MMH rather than RMH), especially if there is a way to draw a boundary between them. An example would be given by two systems connected via a semipermeable membrane that only allows particles of certain types to pass. Quantum tunneling may also play a role in exchange processes. For example, two quantum-mechanical black holes can be in thermal contact (and also in thermal equilibrium as described by the zeroth law) by exchanging Hawking radiation \citep[p. 64]{Wallace2018}. This radiation, which is emitted by black holes, has been attributed to particles tunneling across the hole's event horizon \citep{ParikhW2000}. Again, our ontology should include here both the individual black holes and the joint system.

Moderate compositionalism based on contact is often dismissed based on the fact that it implies the existence of strange objects. For example, if two persons shake hands, a new object would come into existence that ceases to exist after they stop\footnote{The shaking-hands objection against contact as an answer to the SCQ \citetext{\citealp[p. 28]{vanInwagen1987}; \citealp[pp. 35]{vanInwagen1995}} should not be confused with the shaking-hands objection against fastening as an answer to the SCQ \citetext{\citealp[p. 31]{vanInwagen1987}; \citealp[pp. 57--58]{vanInwagen1995}}, which is based on the thought experiment that the two persons shaking hands suddenly become paralyzed and are unable to stop. This is a different argument against a position (fastening) that is not considered in this work.} \citep[p. 35]{vanInwagen1995}. This argument is a form of \ZT{intuitive metaphysics}, as it infers that the contact criterion is wrong from the fact that it has counterintuitive consequences. However, such arguments are not particularly strong - maybe metaphysics just is counterintuitive. The TCP provides a good reason why our intuitions are misleading: van Inwagen's \ZT{shaking hands} objection is based on the assumption that there is no (metaphysically relevant) change if two objects start or cease to be in contact. This, however, is not the case: If two objects start to be in thermal contact, the minus first law implies that they, as a composite system, have a new property - the unique equilibrium state the composite system will approach - that the individual systems did \textit{not} have in the same way.

A different but related objection\footnote{\citet[p. 71]{Silva2013} distinguishes here between the \textit{odd object problem} (an answer to the SCQ implies the existence of strange objects) and the \textit{ordinary object problem} (an answer to the SCQ implies the no-existence of familiar objects).} is that contact implies the non-existence of familiar objects: If composite objects only exist when their constituents are in contact, then, for example, the solar system would not exist \citep[p. 71]{Silva2013}, despite the fact that it is common practice in physics and in everyday language to talk about \ZT{the solar system} and that its existence can be inferred from other plausible criteria such as physical bonding \citep[p. 92]{HusmannN2018}. This problem is easily solved, since it only applies if we consider contact to be a \textit{necessary} condition for composition, such that objects that are not or cease to be in contact do no longer exist. Here, I propose contact as a \textit{sufficient} condition for composition: If two objects are in thermal contact, they form a joint system as required by thermodynamics. This does not exclude the existence of composite systems whose constituents are not in contact (such as \ZT{the solar system}), but which may satisfy other sufficient conditions for composition such as physical bonding. Moreover, the parts of the solar system certainly are able to exchange energy (for example, planets are heated up by the sun), such that if we use \textit{thermal} contact as a criterion for composition, the solar system certainly exists.

I now turn to metaphysical implications. The TCP provides a condition for the $x$s to compose (thermal contact), and is thus a form of moderate compositionalism. However, since the TCP is a sufficient and not a necessary condition for composition, it is also compatible with universalism, as well as with other conditions for composition proposed in the literature, such as bonding \citep{McKenzieM2017,HusmannN2018} or forming a life \citep{vanInwagen1995}. Let us take a look at the practical implications of the TCP. It certainly implies the existence of a variety of interesting objects, including most systems consisting of objects that are in geometrical contact for a time that is sufficiently long to allow for heat exchange (as opposed to, say, two stones that briefly touch each other while flying through the air), as well as many objects that are part of our \ZT{everyday ontology} (such as a block of iron whose parts, if at different temperatures, will exchange heat). In fact, it should be noted that \ZT{being able to exchange conserved quantities} is a very generous condition: In practice, almost all objects will be able to do this in some direct or indirect way. For example, there is a certain gravitational interaction between the Eiffel tower and you underwear, such that they, strictly speaking, satisfy the composition criterion of the TCP. On the other hand, this interaction is far too small to be measurable in practice, and no practitioner of thermodynamics would say that the Eiffel tower and your underwear are in thermal contact.

However, it is interesting here to look at larger scales. It is clearly reasonable to say that the earth (including all objects on its surface) is in thermal contact with the sun. This includes also the Eiffel tower, which will be warmer on a sunny day as it receives heat from the sun. The same will hold, e.g., for your house. Consequently, by the transitivity of thermal contact, the Eiffel tower and your house are, strictly speaking, in thermal contact. Of course, given that the sun is an incredibly hot fusion reactor, the fact that your house is in some sense in thermal contact with the Eiffel tower is of absolutely no relevance compared to the fact that it is in thermal contact with the sun. However, if the Eiffel tower would suddenly have a temperature comparable to that of the sun, we would start to see a significant heat exchange between your house and the Eiffel tower (leaving aside the other consequences this would have). Consequently, the TCP as introduced here would imply the existence of the rather strange object that is the mereological sum of the Eiffel tower and your house. Actually, due to the generality of this argument, it might even imply universalism. This is a remarkable consequence, since universalism is typically motivated based on theoretical considerations and not empirically. However, this \ZT{universalism} might have cosmological restrictions, see section (\ref{funda}) for a discussion. 

This allows us to address an important final point - namely, whether the TCP gives an answer to the SCQ (for macroscopic objects). Interestingly, this is also a matter of contingent facts. Typically, the SCQ is understood as asking for the \ZT{necessary and jointly sufficient conditions} \citep[p. 191]{CarrollM2010} that have to be satisfied by the $x$s such that they compose something. A (valid) objection against the claim that the TCP answers the SQC would therefore be that it merely provides a sufficient condition for composition and does not solve the more difficult problem of giving a necessary condition, which is why I have often referred to it as a \ZT{partial answer}. Three points can be made in response to this objection. First, scientific approaches to the SCQ that infer the existence of composition from the fact that composition is required by a certain scientific theory will generally not be able to provide necessary conditions since they cannot exclude that another scientific theory will require further composite objects. This does not imply that one does not gain any insight from scientific theories - for example, the discussion of quantum entanglement or physical bonding has certainly improved our understanding of the metaphysics of composition. Second, if we succeed in finding a sufficient condition for composition, we have made some progress on the way towards answering the SCQ since we can then exclude nihilism (which states that there are no sufficient conditions for composition). Third, in this specific case a sufficient condition \textit{can} be sufficient as an answer for the SCQ since, as discussed above, it provides an argument for universalism (which is an answer to the SCQ). The reason is that, if we have found a sufficient condition that is satisfied by all $x$s that exist in the actual universe\footnote{Here, I mean the physical universe and not the universe as it is defined in mereology (see section (\ref{compo})).}, we have answered the SCQ at least for our universe (though maybe not for other possible worlds in which some $x$s do not satisfy this condition).

I should also mention an important assumption of the proposed approach: I have inferred the existence of composite thermodynamic systems from the fact that they are an essential part of thermodynamics, for which I have to make the assumption that entities whose existence is assumed in successful scientific descriptions of the world actually exist. This assumption has been challenged by \citet{Brenner2018} in the context of the SCQ, who argued that nihilistic alternatives to scientific theories involving composite objects are not usually considered (such that one cannot say that the compositional aspects of the theories have been tested empirically), and that composite objects are often introduced in science merely as a matter of a more convenient description. However, as discussed above, composition has a much more central status in thermodynamics than merely that of simplifying the description, such that the empirical success of thermodynamics does give us an argument for the existence of composite objects.

\subsection{\label{funda}Thermodynamics as a fundamental theory?}

There is a further important point that needs to be addressed here: Metaphysics has the aim of describing the world at a fundamental level, and the mereological parthood relation is often considered a metaphysically fundamental relation. Thus, if we accept the program of inductive metaphysics, it is natural to try and answer metaphysical questions such as the SQC based on fundamental physical theories such as quantum mechanics. For thermodynamics, however, there are good reasons for not viewing it as a fundamental theory. It is typically introduced as a phenomenological approach applying to macroscopic equilibrium systems whose predictions can be verified from a microscopic derivation only with certain approximations and relative to certain timescales. Thus, if we try to answer a metaphysical question using thermodynamics, we might be, in the words of \citet{Callender2001}, \ZT{taking thermodynamics too seriously}.

However, as discussed in more detail by \citet[p. 397]{Needham2013}, there is no reason not to take a scientific theory metaphysically seriously just because it is not \ZT{fundamental}, since, even if it can be reduced to a more microscopic theory, it may still make correct (ontological) claims about the world. In particular, approaches the SQC based on non-fundamental scientific theories are not uncommon. For example, \citet{HusmannN2018} argue for the existence of the solar system as a composite entity based on the criterion of physical (gravitational) bonding - this argument is based on Newtonian mechanics. Moreover, \citet{Brenner2018} discusses (though not endorses) the idea that evolutionary biology requires us to see species as mereological sums (see \citet{Hull1980}). Whatever our opinion on such ideas might be, we should not dismiss an otherwise valid argument for composition based on natural selection simply because its premises do not involve an appeal to the standard model of particle physics. Rather, we should pay attention to the fact that the validity of the metaphysical argument is limited by the range of validity of the scientific theory that our argument is based on. For example, if one succeeds in showing that natural selection provides an argument for viewing a species as a composite entity in its own rights, then it is reasonable to believe that all the tigers in the world (assuming tigers exist) compose an entity \textit{Panthera tigris}. Such an argument would, however, not provide any information regarding the question whether stars compose a galaxy, simply due to the fact that the theory of natural selection does not apply to astrophysical systems. This is what distinguishes answers to the SQC based on the special sciences\footnote{I understand the term \ZT{special science} in the same way as \citet[p. 195]{RossLC2007}, i.e., as referring to a science that has a restricted domain of application. A different terminology would be to define \ZT{special science} as \ZT{all sciences other than physics}.} from those based on (presumably) fundamental theories such as quantum field theory.

The TCP is thus applicable (only) to those systems which the laws of thermodynamics can be applied to. I now illustrate what this implies for the applicability of the TCP using a few examples:
\begin{enumerate}
    \item A nonequilibrium system.
    \item Two individual (quantum-mechanical) particles.
    \item Two stars at opposite ends of the universe.
    \item A self-gravitating system.
\end{enumerate}
For (1), the situation is rather clear. As emphasized above, my argument rests on the applicability of the minus first law, stating that a system approaches a unique equilibrium state. The minus first law - unlike the other axioms of thermodynamics - applies also to nonequilibrium systems, since it is primarily concerned with the approach to the equilibrium state (rather than simply with the properties of equilibrium systems). For example, \citet[p. 139]{teVrugt2020} has interpreted the H-theorem of dynamical density functional theory \citep{Munakata1994,teVrugtLW2020}, stating the monotonous decrease of the free energy functional (a clear nonequilibrium result), as a manifestation of the minus first law of thermodynamics. Things get slightly more complicated for driven and active systems (see \citet{Menzel2015}), which not only are not in equilibrium, but also do not approach it. Here, one has to consider the equilibrium state the system would approach if it would not be driven out of equilibrium.

Regarding (2), the answer depends on whether thermodynamics can be applied to small quantum systems. This, in turn, depends on what we mean by \ZT{thermodynamics}. In philosophy, it is common to make a sharp distinction between \ZT{thermodynamics}, a macroscopic theory based on phenomenological axioms, and \ZT{statistical mechanics}, which is based on a microscopic description (see, e.g., \citet[p. 309]{Uffink2001}, \citet[p. 72]{Brown2005}, and \citet[p. 139]{teVrugt2020}). Paying attention to this difference is important, e.g., when studying problems of intertheoretical reduction. For our purposes, however, this distinction is not relevant, such that we can resort to the more blurry terminology of physicists, who use the word \ZT{thermodynamics} also for microscopic theories. What is relevant for our purposes is not whether our statements belong to the theory of classical thermodynamics, but simply on whether they are true for the objects we are interested in. For small quantum systems, much progress has been made here in recent years in the field of \textit{quantum thermodynamics} \citep{VinjanampathyA2016,GemmerMM2009}, which aims to extend conventional thermodynamics to small systems in which quantum effects are relevant.

While technical modifications are certainly required here, the essential concepts required for the mereological argument might still go through. For example, the concept of thermal contact and heat flow can be extended to qubits prepared in thermal Gibbs states \citep{MicadeiEtAl2019,JevticJR2012}. Experiments on heat flow in quantum systems between $^{13}$C and $^1$H nuclei have been performed by \citet{MicadeiEtAl2019}, showing that heat flows from the hot to the cold spin if they are initially uncorrelated. Similarly, the equilibration of a system of two atomic nuclei in thermal contact was studied by \citet{SchmidtJ2011}. These can exchange not only energy but also nucleons, with the latter being the dominant mechanism for energy transport. The entropy of the two-nucleon-system evolves towards its maximum value in agreement with the laws of thermodynamics. Finally, in dissipative spontaneous collapse models \citep{SmirneVB2014}, even single-particle quantum systems can be shown to obey the dynamics typically associated with thermal equilibration \citep{teVrugtTW2021}.

Note that quantum mechanics poses additional mereological problems, which have been discussed by \citet{NaegerS2020} and \citet{Naeger2020}. For instance, if the system of two nuclei is in an entangled state, one might question whether we are dealing with one or two objects. As shown by \citet{Naeger2020}, there are good reasons to believe in the existence of two separate objects also in this case, since the individual objects carry fundamental properties (such as their mass and charge) that cannot be reduced to macroscopic properties. If the two nuclei are placed in thermal contact, the existence of the composite system then follows from the TCP by the line of argument presented above.

The experiments on two spins in thermal contact by \citet{MicadeiEtAl2019} are interesting also in another regard: Given that the microscopic laws of physics are invariant under time-reversal, it requires an explanation that the thermodynamic arrow of time always has the same direction. Typically, this is attributed to the choice of initial conditions, an idea known as the \ZT{past hypothesis} \citep{Wallace2011}. Anti-thermodynamic behavior is never observed because this would require extremely fine-tuned initial conditions \citep{Popper1956}. This suggests that it is possible, in principle, to achieve such behavior by setting up a system in such an initial condition. While this is impossible for macroscopic systems, it can be doable for small ones: \citet{MicadeiEtAl2019} observed that, if the two spins are initially correlated, spontaneous heat flow from the colder to the hotter spin does occur. While this certainly has interesting consequences for thermodynamics, the TCP can be justified also for this case. Recall that the justification of the TCP is based on the fact that there exists a \ZT{special state} (the equilibrium state) that is realized by the combined system. Such a state still exists here, it is just not the \textit{final} (equilibrium) state, but the (highly correlated!) \textit{initial} state. See section (\ref{lir}) for a further discussion of this point.

Next, we turn to system (3), consisting of two stars that are extremely far apart. Due to their distance, heat flow of the kind considered in equilibrium thermodynamics is impossible. Consequently, they are not in thermal contact. This, however, does not imply that the TCP is not applicable, it simply means that, according to the TCP, these two stars do not compose an object. Hence, these two stars do not pose a problem for the TCP. They might, however, pose a problem for the claim made above that the TCP provides an inductive argument for universalism, given that we have just found two objects that are not in thermal contact and thus do not (necessarily) compose. However, this claim is not lost entirely: Recall that the argument for the existence of a composite object consisting of different systems in thermal contact was based on the fact that this composite object converges to a unique equilibrium state. Thus, the TCP forms an inductive basis for universalism if the universe as a whole can be expected to approach equilibrium. The idea of an equilibration of the universe as a necessary consequence of thermodynamic irreversibility arose in the 19th century \citep{Kelvin1862} and is known as the \ZT{heat death}, which remains an important cosmological hypothesis \citep{Frautschi1982}. It is, however, an open question whether (something like) a heat death will actually be the fate of the universe. On the one hand, this depends on the origin of thermodynamic irreversibility in general - for interventionists, who explain irreversibility by external perturbations, the entropy of the universe as a whole (assuming that it is a closed system) has to be constant \citep[pp. 1261-1262]{RidderbosR1998}. On the other hand, modern cosmology might not support this hypothesis. For example, \citet{Frautschi1982} has argued that in an expanding universe, although the entropy will keep growing (even at late times through events such as the collapse of galaxies and galaxy superclusters to black holes, quantum tunneling of nuclei to iron, or quantum tunneling of matter into black holes), the \textit{ratio} of the entropy to its maximum will decrease. This would imply that the universe, while continuously increasing its entropy, would not reach equilibrium. Nevertheless, \citet[p. 369]{AdamsL1997} suggest that it is still possible that the universe reaches a state of purely adiabatic expansion, making it \ZT{a dull and lifeless place with no ability to do physical work}. Whether the TCP actually supports universalism thus depends on the future development of cosmology.

Finally, the case of a self-gravitating system (4) poses another astrophysical problem. This is a consequence of the \textit{gravothermal catastrophe} \citep[pp. 523-525]{Wallace2010}: Gravity-dominated systems have negative heat capacity, i.e., their temperature increases if they emit heat. Thus, if a hot self-gravitating system is in contact with a colder environment, the initial temperature difference will increase rather than decrease. In astronomy, effects of this form are known to arise in globular clusters (dense spherical systems of stars). Thus, Newtonian gravity-dominated systems will, in general, not reach a stable equilibrium state \citep[p. 55]{Wallace2018}. This poses a problem for the TCP since it relies on the fact that systems reach equilibrium, suggesting that it does not apply to self-gravitating systems. However, there do exist gravity-dominated systems at thermal equilibrium, such as neutron stars and black holes. Stars are stabilized by fusion reactions, which are possible as long as sufficient fuel is available \citep[p. 55]{Wallace2018}. If the matter of the star contracts (and if its mass is sufficiently large), electrons and protons will combine to neutrons, leading to a neutron star (which has positive heat capacity). At even larger masses, the system will collapse to a black hole, which has negative heat capacity and (as is well known) a general tendency to absorb everything in its environment \citep[pp.526-527]{Wallace2010}. For black holes, a theory of equilibrium thermodynamics exists, and thermal contact between black holes and other objects is possible once Hawking radiation is taken into account. This is discussed in detail in \citet{Wallace2018} (see also \citet{Wallace2019}).

In summary, I have shown that, although \textit{classical} thermodynamics is a phenomenological theory for medium-sized objects, its principles hold (perhaps in a generalized form) in a large variety of contexts, including small systems from nuclear physics and large astrophysical objects. Consequently, the TCP has far-reaching implications for the problem of composition. Moreover, I have explained that even if a scientific theory such as thermodynamics is not universally applicable, it can still deliver valuable metaphysical insights for the domain where it is.

\subsection{\label{lir}Structural realism and logic in reality}
In addition to the question addressed in section (\ref{funda}), i.e., whether inferring mereological inferences from thermodynamics is justified, one might ask the even more basic question of whether looking for mereological principles is a reasonable project of metaphysics in general. This issue needs to be addressed because it is controversial: In their book \textit{Every Thing Must Go}\footnote{A note on citations: Since the different chapters of \citet{LadymanR2007} have different authors, they are listed as separate entries in the bibliography.}, \citet{LadymanR2007} argue that metaphysical mereology is based on an outdated picture of the world as being a \ZT{container} \citep[p. 3]{RossLS2007} containing objects that are themselves containers. In this context, typical metaphysical approaches to mereology, such as those by \citet{vanInwagen1995} or \citet{Markosian2005}, are accused of ignoring basic results of fundamental physics. Instead, \citet{LadymanR2007} argue that fundamental physics - in particular quantum mechanics - allows to make a case for the view that the world is fundamentally based on \textit{structure}, which is real and not supervening on individual objects in the classical sense (\ZT{ontic structural realism}). The view that classical mereology is not applicable to quantum systems is not shared universally (see, e.g., \citet{Naeger2020} for a different position which assumes that classical mereology can be applied to quantum systems under certain symmetry assumptions). Nevertheless, given that I am explicitly committed to the project of an \textit{inductive} metaphysics that is informed by science (here: thermodynamics), the position that modern physics poses a threat for the metaphysical project of analyzing composition relations deserves to be taken seriously.

Various things can be said in response to this position. First, \citet[p. 21]{RossLS2007} explicitly mention that various special sciences are concerned with composition, for example when biologists investigate how a multicellular organism is composed of cells. Their objection is that \ZT{Metaphysicians do not dirty their hands with such details, but seek instead to understand [...] the general composition relation itself}, and that this hypothetical general composition relation is just \ZT{an entrenched philosophical fetish}\citep[p. 21]{RossLS2007}. In the present work, I have precisely the aim of dirtying my hands with composition relations (or better: parthood relations) in a special science, namely thermodynamics. If we assume that what is true scientifically should also be true metaphysically, then it is safe to argue that thermodynamic systems can be composed of other thermodynamic systems. What is a different matter is whether the claim by \citet[p. 21]{RossLS2007} that there are only composition relations sui generis from the special sciences, but not a fundamental \ZT{metaphysical} type of composition, is correct. However, the SCQ can be meaningful even in this case if it is rephrased appropriately, for example as \ZT{What are the necessary and sufficient conditions for some $x$s to compose according to a particular form of composition?} In this work, we would then be concerned with \textit{thermodynamic} composition.

Second, the problem is of course that the discussion is assuming the world to consist of separate objects with clear identity that stand in parthood relations. This is a consequence of the origin of mereology in classical predicate logic (see section (\ref{compo})). If one, in the tradition of ontic structural realism, denies that the world is composed of such objects, the question remains how to make sense of statements of the form \ZT{This system is composed of two boxes in thermal contact}. \citet{RossLC2007} discuss two options for dealing with the fact that individual objects play a more central role in the special sciences than they do in quantum physics. First, one can deny individualism for fundamental physics but hold on to it for the special sciences. Second, one can understand the entities that the special sciences are concerned with as \ZT{real patterns}. (\citet{LadymanR2007} endorse the second option, I refer to their treatment for a discussion of what a \ZT{real pattern} is.) For our purposes, both options are promising. Regarding option one, we then need to address whether the entities we are concerned with deserve the status of objects. When it comes to the ones discussed in section (\ref{funda}), this might in particular be disputed for atomic nuclei (quantum objects) and for black holes. (See \citet[pp. 56-57]{Wallace2018} for a discussion of whether black holes qualify as objects.) Regarding option two, one would then, strictly speaking, have to replace \ZT{object} by \ZT{real pattern} in all statements I make. Nevertheless, these considerations show that the TCP (perhaps in a modified form) can be defended even from a position of ontic structural realism.

If the aforementioned problems of mereology result from the fact that it arises from classical logic, concerned with separable objects and their properties, it is worth taking a look at non-classical forms of logic in general and mereology in particular. This might then again provide us with an understanding of parthood relations that is universal as opposed to depending on particular special sciences. Such an approach is given by \textit{logic in reality} (LIR), which was developed by \citet{Brenner2008} based on ideas by the Romanian philosopher St\'{e}phane \citet{Lupasco1951} (see \citet{Brenner2010}). LIR is a non-classical non-propositional logic based on the \textit{principle of dynamic opposition} (PDO), according to which all logical elements (corresponding to objects, processes, or events) evolve non-reflexively between actualization and potentialization of themselves and their opposite. In other words, rather than having propositions that are either true or false, one has logical elements \textbf{e} that have a certain degree of actualization and a \ZT{contradiction} non-\textbf{e} that is potentialized to the extent that \textbf{e} is actualized and vice versa. The actualization and potentialization can never reach 100\% (asymptoticity). The point of maximal contradiction, at which \textbf{e} and non-\textbf{e} have the same degree of actualization, is called \ZT{T-state}. Consequently, LIR is a formalism with an included middle \citep[pp. 1-17]{Brenner2008}. This general principle has also been applied to mereology \citep[pp. 128-129]{Brenner2008}. Here, the relation between a whole and is parts is assumed to be dynamic. Parts and whole are not independent from each other \citep[p. 413]{BrennerI2021}. Instead, aspects of the parts can be potentialized in the whole and vice versa.

\citet[p. 101]{Brenner2008} points out that LIR, by not viewing parts and wholes as separable in the way a classical theory of individuals does, is consistent with the ontic structural realism of \citet{LadymanR2007}. In fact, LIR has close connection to (and is influenced by) quantum mechanics. Non-separability, which is a key feature of entangled quantum systems, is incorporated in LIR and attributed there to the failure of classical individuation \citep[p. 101]{Brenner2008}. The mereology of LIR allows to capture the impossibility of a bi-directional reduction between parts and wholes by the dynamic relation between them implied by the PDO \citep{BishopB2017}. In this work, we are not (primarily) concerned with quantum entanglement. However, as shown by \citet{AertsDG2000}, logical principles similar to those of quantum mechanics can also apply on macroscopic levels. An example are cognitive processes \citep{AertsGS2013}.

The mereology of LIR can also be helpful for understanding certain aspects of the TCP. Recall that, as discussed above, the TCP implies composition also for non-equilibrium systems in thermal contact. The reason is that such a non-equilibrium system still has a unique equilibrium state it spontaneously approaches, and this equilibrium state is a property of the \textit{combined} system. It might now be objected that the system is not in this state (although it approaches it), such that this state cannot lead to composition at the present time. From the LIR point of view, we can understand the approach to equilibrium as a process of actualization of the equilibrium state. When two systems with different temperatures are put in thermal contact - leading to a non-equilibrium initial state - the equilibrium state that the system subsequently approaches is \textit{potentialized} in the parts of the system. Later, this state, which is a property of the whole, is \textit{actualized} in the process of equilibration.\footnote{The idea of an actualized/potentialized equilibrium state can also be found in the LIR analysis of signal transmission in nerve cells, where (electrostatic) equilibrium is first actualized and then potentialized by an excitation \citep[p. 314]{Brenner2008}.}

In particular, such ideas can help to incorporate the case of \ZT{reversed heat flow} in the experiments by \citet{MicadeiEtAl2019} discussed in section (\ref{funda}). There, heat flow from a cold to a hot spin was observed when the two spins were initially correlated, reflecting the fact that the thermodynamic asymmetry of time is a consequence of initial conditions.\footnote{This assumes that the laws of physics themselves are time-symmetric. An alternative explanation of thermodynamic equilibration based on a time-asymmetric modification of quantum mechanics has been suggested by \citet{Albert1994b} (see \citet{teVrugtTW2021} for a discussion).} While this can put limits to the validity of the minus first law, the TCP can still be defended based on the fact that the system still potentializes a unique equilibrium state. The difference to the standard case without fine-tuned initial conditions is \ZT{only} the fact that thermodynamic equilibrium gets potentialized rather than actualized during the time evolution of the system.

At this point, it is interesting to note that a connection between thermodynamics and LIR was established already in \citet{Brenner2008}, \citet{Brenner2012}, and \citet{BrennerI2021}, where the second law of thermodynamics was understood as expressing a universal tendency towards homogenization. Here, I mention two further possible ways in which such a connecton can be established: First, quantitative models of equilibration show that the relaxation to equilibrium is an asymptotic process, since (as can be seen from a linear stability analysis \citep[pp. 198-199]{teVrugtLW2020}) it is exponential and thus requires (in principle) an infinitely long time. This can be thought of as a manifestation of the asymptoticity principle of LIR. Second, events in LIR do not take place in time. Instead, time is \ZT{unrolled} by the dynamics of potentialization or actualization. A state of complete actualization/potentialization would correspond to a non-dynamic state of complete equilibrium  \citep[pp. 231-232]{Brenner2008}. This has an interesting connection to the idea expressed by \citet{Schrodinger1950} and \citet{Hawking1994} that the second law of thermodynamics might be true by definition in the sense that the future is \textit{defined} as the direction of time in which entropy increases. (See \citet[pp. 532-535]{BrownU2001} for a critical discussion of such approaches.) From the LIR point of view, such a determination of time by thermodynamic processes would be a manifestation of the fact that time (here: the thermodynamic arrow of time) depends on the dynamics of potentialization/actualization (here: the process of homogenization that is thermal equilibration).\footnote{See \citet[p. 141]{BrennerI2021} for a discussion of \ZT{entropy time}.}

\section{\label{compo}Composition in thermodynamics - a logical perspective}
Up to now, the discussion has been rather informal. Based on physical and metaphysical arguments, I have argued that objects in thermal contact compose a further object (thermodynamic composition principle). Next, I will analyze how this principle can be stated in a more formal way using the logical method of mereotopology.

\subsection{\label{mtop}Mereotopology}

From a logical perspective, mereology is an extension of first-order predicate logic with identity. Early work on this topic was done by \citet{Husserl1900} and \citet{Lesniewski1916}. A careful discussion of the formalism can be found in \citet{Hovda2009}, a historical overview in \citet{GruszczyV2015}, and a general presentation in the handbook by \citet{BurkhardtEtAl2017}. Mereology can be extended further by combining it with ideas from topology. The resulting formalism is known as \textit{mereotopology} and allows, e.g, to discuss concepts such as connectedness and boundaries \citep{Smith1996}. This will be useful for our discussion of thermodynamics. I here present a mereotopological formalism in which \ZT{being connected} ($C$) is a primitive predicate (as done, e.g., by \citet{Varzi1996}). Alternatively, one could, following \citet{Smith1996}, introduce a primitive predicate \ZT{interior part} and then define the connection in terms of it. However, this would make my presentation less compact, since I here only require connectedness.

I start by introducing mereology. It adds to predicate logic a relation $<$, whose intended interpretation is that $x < y$ stands for \ZT{$x$ is a part of $y$}\footnote{On a purely formal level, $<$ is of course just a symbol with certain properties that in other contexts may also have other intepretations such as \ZT{subsethood} (see below).}. It has the properties \citep[p. 63]{CalosiT2014}
\begin{align}
    &x < x \text{ (Reflexivity)},\label{reflexivity}\\
    &x < y \land y < x \rightarrow x=y \text{ (Anti-symmetry)},\label{antisymmetry}\\
    &x < y \land y < z \rightarrow x < z \text{ (Transitivity).\label{transitivity1}}
\end{align}
These are the properties one would intuitively expect a parthood relation to have. Based on the predicate $<$, we can define the relations \citep[p. 261]{Varzi1996}
\textit{overlap} (i.e., sharing parts)
\begin{equation}
Oxy \define \exists z(z < x \land z < y)
\end{equation}
and \textit{underlap} (i.e., being part of the same thing)
\begin{equation}
Uxy \define \exists z(x < z \land y < z).\label{underlap}
\end{equation}
The axioms (\ref{reflexivity}), (\ref{antisymmetry}) and (\ref{transitivity1}) constitute a minimal theory known as \textit{ground mereology} (M) \citep[p. 260]{Varzi1996}. They introduce parthood as a partial order. An extension is \textit{extensional mereology} (EM), which also includes the \textit{strong supplementation principle} \citep[p. 262]{Varzi1996}
\begin{equation}
\neg (y < x) \rightarrow \exists z (z < y \land \neg Ozx).    
\end{equation}
Adding the further axioms \citep[p. 43]{CasatiV1999}
\begin{align}
&Uxy \rightarrow \exists z \forall w (Owz \leftrightarrow (Owx \lor Owy)),\label{sum}\\
&Oxy \rightarrow \exists z \forall w (w < z \leftrightarrow (w < x \land w < y))
\end{align}
gives \textit{closed extensional mereology} (CEM). The axiom (\ref{sum}) is of particular interest because it introduces the idea of a \textit{mereological sum}: If $x$ and $y$ are part of the same thing, there is an object $z$ of which they are both parts and which has no parts that do not overlap with $x$ or $y$. This gives for two underlapping objects a minimal underlapper (mereological sum). I write the mereological sum of two objects $x$ and $y$ as $x+y$. Using a description operator $\tau$, it can be defined as \citep[p. 263]{Varzi1996}
\begin{equation}
x+y := \tau z \forall w (Owz \leftrightarrow (Owx \lor Owy)).
\end{equation}
A well-known example satisfying the axioms of CEM is set theory with the subsethood relation $\subseteq$. In this case, the mereological sum of two sets $a$ and $b$ would be the union $a \cup b$. Note, however, that subsethood is just \textit{one} possible interpretation of the $<$ relation. In general, mereology does not require that the objects we are talking about are sets. Historically, the wish for a nominalist alternative to set theory was an important motivation for the development of mereology \citep[p. 414-415]{GruszczyV2015}. \footnote{A detailed discussion of set theory in a mereological context can be found in \citet{Lewis1991}.} 

What mereology does not offer is an answer to the question whether objects are connected (this concerns both two different objects and the parts of one object). A good example is a bikini, which consists of two disconnected parts \citep[p. 4]{Rachavelpula2017} and is thus a \textit{scattered object}. It is here where topology comes into play\footnote{The mereological approach to topology considered here differs from the set-theoretical way topology is usually introduced in mathematics lectures. There is, however, a connection between mereotopology and the study of topological spaces in mathematics, see \citet{Rachavelpula2017} for a discussion.}: In \textit{ground topology} (T), one introduces a predicate $C$ with the intended interpretation \ZT{is connected to} that satisfies the axioms \citep[p. 268]{Varzi1996}
\begin{align}
&Cxx,\label{gt1}\\
&Cxy \rightarrow Cyx,\label{gt2}\\
&x < y \rightarrow \forall z (Czx \rightarrow Czy)\label{gt3}.
\end{align}
Here, (\ref{gt1}) states that everything is connected to itself, (\ref{gt2}) makes $C$ a symmetric relation, and (\ref{gt3}) ensures that if something is connected to a part of an object, it is connected to the object (monotonicity with respect to parthood \citep[p. 6]{Rachavelpula2017}). One can now combine the axioms of T with those of some form of mereology. For example, adding the axioms of T to the axioms of M gives MT, while CEMT is constructed by combining T with CEM. CEMT is a powerful framework that allows, e.g., to introduce the idea of a \textit{self-connected} object in the form
\begin{equation}
SCz \define \forall x \forall y (z = x+y \rightarrow Cxy),
\label{self}
\end{equation}
which formalizes the idea that one cannot divide an object into two parts that are disconnected \citep[p. 271]{Varzi1996}. Mereologically, this is interesting because it allows to distinguish a connected whole from a scattered object, since scattered objects are not self-connected. (The definition (\ref{self}) uses the idea of a mereological sum and therefore has to be changed in MT, this point is discussed in section (\ref{selfnonself}).)

\subsection{Axiomatization of the TCP}

Having established that thermodynamics requires multiple systems in thermal contact to compose a single system, I now turn to the consequences of this result for the axioms of mereotopology. Thereby, I will show that a logical formalism incorporating the TCP is given by CEMT together with the principle \citep[p. 58]{CasatiV1999}
\begin{equation}
Cxy \rightarrow Uxy.   
\label{gt4}
\end{equation}
The extension of CEMT obtained by adding (\ref{gt4}) will be referred to as CEMT$^\star$. Note that my argument does not imply that there cannot be a stronger formal system, e.g., one that involves unrestricted composition or at least additional sufficient conditions for the existence of a mereological sum. I merely argue that thermodynamics provides an argument for these two principles. This is done in three steps. First, I motivate the choice of CEMT$^\star$. Second, I show that weaker systems are not sufficient. Third, I discuss possible counter-arguments.

For the first step, recall that the TCP states that systems in thermal contact compose an object. Since CEM implies that any two underlapping objects have a sum \citep[p. 43]{CasatiV1999}, it is sufficient if systems in thermal contact underlap. Formally, the TCP then requires that
\begin{equation}
TC xy \rightarrow Uxy.
\label{trivial}
\end{equation}
We thus need to formalize the relation $TC$ (being in thermal contact) in such a way that (\ref{trivial}) is satisfied. (Of course, we could simply introduce $TC$ as a primitive relation satisfying (\ref{trivial}), but this would not gives us many formal insights into the mereological consequences of thermodynamics.) A useful starting point is the contact relation $C$, which I will, from now on, use to denote contact with the possibility of direct exchange of heat (or other conserved quantities).\footnote{This means that $C$ does not denote geometrical contact, since it is possible that two systems that do not touch each other exchange heat by radiation. Nevertheless, if we think of radiation as a connection between two objects and ignore the theoretical case of perfect thermal isolation, the interpretation of $C$ used here will essentially coincide with the usual interpretation.} This is compatible with the axioms (\ref{gt1}) - (\ref{gt3}). Moreover, we have to take into account that, e.g., heat exchange is also possible in an indirect way: If two objects $a$ and $b$ are not in contact with each other, but are both in contact with a third object $c$, they might exchange heat indirectly via $c$. This is why, as required by the zeroth law, thermal contact is (in contrast to contact) a transitive relation. If we interpret \ZT{connected to} ($C$) in the way proposed here, two objects $x$ and $y$ can exchange conserved quantities if they are connected or if there are objects connecting them. In both cases, we can consider the mereological sum of $x$, $y$ and (if required) the further objects connecting them. This sum (let us call it $z$) will then be a self-connected object as defined by (\ref{self}), since, if we split it into two parts, the parts can exchange conserved quantities and are thus connected. Therefore, we can define the relation $TC$ as
\begin{equation}
TCxy \define \exists z (SC z \land (x < z) \land (y < z)).
\label{tc}
\end{equation}
Informally, (\ref{tc}) states that $x$ and $y$ are in thermal contact if there is a self-connected object they are both part of. In this case, they will be able to exchange conserved quantities either directly or indirectly. If we define $TC$ by (\ref{tc}), the principle (\ref{trivial}) will automatically be satisfied: $TCxy$ implies by (\ref{tc}) the existence of an object $z$ that $x$ and $y$ are both part of, which by (\ref{underlap}) implies that they underlap. 

However, the line of argument that has led to the definition (\ref{tc}) assumes that the mereological sum $z$ exists, which is only guaranteed if (self-connected) objects that are in contact add up to a self-connected whole. As discussed by \citet{CasatiV1999}, this is ensured by CEMT$^\star$. First, CEM(T) implies that if two objects underlap, they have a unique mereological sum. From (\ref{gt4}), we can infer that objects that are in contact do underlap. Consequently, the mereological sum of two objects in contact exists. In particular, CEMT$^\star$ allows to show that \citep[p. 58]{CasatiV1999}
\begin{equation}
(SCx \land SCy \land Cxy) \rightarrow SCx+y.
\label{connectedsum}
\end{equation}

Moreover, since it follows from the definition of $E_2xy$ that $E_2 xy \rightarrow TCxy$ (see section (\ref{laws})), the zeroth law of thermodynamics (\ref{zerothlaw}) requires that $TC$ is a transitive relation. This is, in CEMT$^\star$, ensured by the definition (\ref{tc}): If $TCxy$ and $TCyz$ hold, there is a system $u$ with $x < u$, $y < u$ and $SCu$, and a system $v$ with $y < v$, $z < v$ and $SCv$. We have $Cyy$ by (\ref{gt1}), which by (\ref{gt3}) implies $Cyu$ since $y < u$. From $Cyu$ and $y < v$, it follows by (\ref{gt3}) that $Cuv$. Consider the system $w = u + v$. It will, by (\ref{connectedsum}), satisfy $SCw$. Moreover, by (\ref{transitivity1}), it satisfies $x < w$ and $z < w$. This implies by (\ref{tc}) that $TCxz$, such that the relation $TC$ defined by (\ref{tc}) is transitive.

Next, I show that axiomatic systems weaker than CEMT$^\star$ are not sufficient\footnote{More precisely: That we require both the existence of mereological sums of underlapping objects and the composition principle (\ref{gt4}).}. Suppose that we only have (\ref{gt4}). In this case, it is not guaranteed that any two objects in thermal contact compose a system. For many mereological models, it is true that
\begin{equation}
\exists z \forall x (x < z)\label{universe}.
\end{equation}
Informally, (\ref{universe}) states that there is something ($z$) that everything is a part of. If this $z$ is unique (which it is  if we have extensionality), it is called \ZT{the universe}. When (\ref{universe}) is true, the underlap relation $U$ is trivial \citep[p. 264]{Varzi1996}. In a system that, in addition to MT, only involves (\ref{universe}) as an axiom, (\ref{gt4}) would clearly be satisfied. However, it would be entirely possible that there exists no composite object other than the universe. 

This problem does not arise in CEMT$^\star$: In this case, two underlapping objects are guaranteed to have a minimal underlapper (mereological sum). If two objects that are in contact underlap (by (\ref{gt4})), they will therefore have a sum. Consequently, if we add (\ref{gt4}) to CEMT, we can add up connected objects to a single one. As an example, we take three objects in thermal contact. Let us call them $a$, $b$ and $c$ and assume, without loss of generality, that $Cab$ and $Cbc$ (there has to be a direct or indirect connection between any two of these three objects for thermal contact). If $Cab$ holds, there exists (given CEMT$^\star$) an object $a+b$. From (\ref{gt3}), $Cbc$ and $b < a+b$, we can infer that $c$ is in contact with $a+b$. This then implies that the mereological sum of $a+b$ and $c$ also exists, which is precisely the object we are looking for. The same argument applies if the number of objects in contact is larger.

On the other hand, we also cannot do without (\ref{gt4}). The composition principle (\ref{sum}) of CEM is a conditional principle \citep[p. 68]{CalosiT2014}, i.e., it only guarantees the existence of a mereological sum for objects that underlap. Without a universe, this is not generally the case in CEM. 

\subsection{\label{selfnonself}Self-connected systems that are not self-connected}

Another reason why we require the axioms of CEMT (and not just those of MT) is that, in (\ref{tc}), we have defined the predicate $TC$ (thermal contact) based on the predicate $SC$ (self-connectedness). The idea of self-connectedness is central for the mereotopological approach to thermodynamics, since self-connected objects have an equilibrium state as a joint system, whereas scattered objects would simply be in equilibrium if their constituents are (such that the existence of the joint system is not required by the TCP). In CEMT, $SC$ is defined by (\ref{self}). This definition is based on the idea of a mereological sum and thus is not applicable in MT. Therefore, the definition of $SC$ used in MT is \citep[p. 57]{CasatiV1999}
\begin{equation}
SC x \define \forall y \forall z (\forall w (Owx \leftrightarrow Owy \lor Owz) \rightarrow Cyz).
\label{selfmt}
\end{equation}
Informally, it is (like (\ref{self})) based on the idea that any two objects $y$ and $z$ that form an object $x$ must be in contact for $x$ to be self-connected. The problem is now that (\ref{selfmt}) can also be satisfied for scattered objects, which contradicts the idea that \ZT{being self-connected} is the negation of \ZT{being scattered}. To see this, let our domain of discourse contain objects $a$, $b$, $c$ and $d$ (with $a$, $b$ and $c$ nonoverlapping), such that $d$ is composed of $a$, $b$, and $c$. Moreover, let $a$ and $b$ be in contact ($Cab$), while $c$ is not connected to $a$ or $b$. Then, $d$ is a scattered object. We work in MT, i.e., we do not necessarily have such a thing as a mereological sum of $a$ and $b$ (an object made up of $a$ and $b$ and nothing else), and we assume that there is indeed no object composed of only two of the three objects $a$, $b$ and $c$. Now, we can demonstrate that $d$ satisfies the definition (\ref{selfmt}): For this to be the case, $\forall w (Owx \leftrightarrow Owy \lor Owz)$ must imply $Cyz$ for all $y$ and $z$. An implication can only be false if the antecedent is true and the consequent is not. $Cyz$ is, in our domain of discourse, only wrong if $y$ and $z$ are given by $a$ and $c$ or $b$ and $c$. Without loss of generality, we consider $y=a$ and $z=c$. For $d$ to be not self-connected according to the definition (\ref{selfmt}), it therefore has to be true that $\forall w (Owd \leftrightarrow Owa \lor Owc)$. However, there is at least one $w$ - namely $b$ - which satisfies $Owb$, but not $Owa \lor Owc$. Thus, $d$ is a self-connected object according to (\ref{selfmt}). 

Informally, this problem arises because (\ref{selfmt}) defines an object as being self-connected if any \textit{two} objects composing it are in contact. The object $d$ is, however, composed of \textit{three} objects. In contrast, if we work in CEMT, we can consider the mereological sum $a+b$ to decompose $d$ into the two objects $a+b$ and $c$ (that will then not be connected), such that this problem does not arise for the definition (\ref{self}) in CEMT. In our case, the problem with (\ref{selfmt}) would imply that two objects could satisfy the definition (\ref{tc}) of $TC$ that are not actually in thermal contact. Consequently, if our axioms do not imply the existence of a mereological sum, we cannot define the relation $TC$ based on the predicate $SC$ as done in (\ref{tc}). Moreover, the aforementioned problem of the definition (\ref{selfmt}) shows that MT is, in general, not sufficient to identify scattered objects, which is problematic also beyond applications to thermodynamics. 

\section{\label{conclusion}Conclusion}
In this article, I have shown that the minus first law of thermodynamics, which states that isolated systems spontaneosly approach a unique equilibrium state, suggests a new approach to the special composition question, the \ZT{thermodynamic composition principle} (TCP): Multiple objects that are in (generalized) thermal contact compose an object. This suggestion is based on a systematic analysis of possible mereological models for systems in thermal contact, which shows that the first-order predicate \ZT{being in equilibrium} should be thought of as a property of the joint system rather than as a relation between or a collective property of its individual constituents. On a formal level, the TCP can be incorporated into the mereotopological system CEMT combined with the axiom that systems in contact always underlap. This is done by defining \ZT{thermal contact} based on the topological predicate \ZT{self-connectedness}.

\begin{acknowledgements}
I am very grateful to Paul M. N\"ager for introducing me to the theory of composition and mereology in science, and for detailed feedback on previous versions of this article. Moreover, I thank Niko Strobach for an introduction to the formalism of mereology and its role in metaphysics. Also, I thank Ulrich Krohs, Oliver Robert Scholz, Ansgar Seide and the other participants of the thesis colloquium where a previous version of this article was presented, for helpful comments. I am particularly indebted to Alice Rolf and Wolfram Pohlers for interesting conversations about topology and logic, and to Raphael Wittkowski for discussions on physical aspects. I also thank David Wallace for providing literature. This work was presented at the conference \ZT{'Going up?' Realisation and Composition across the Sciences} hosted by the University of Bristol, I thank the participants for interesting questions and suggestions. Finally, I acknowledge stimulating discussions about scientific metaphysics in the \ZT{Arbeitskreis Wissenschaftstheorie} (Philosophy of science discussion group) at the Center for Philosophy of Science in M\"unster. \\
This work is funded by the Deutsche Forschungsgemeinschaft (DFG, German Research Foundation) -- grant number WI 4170/3-1. I also wish to thank the Studienstiftung des deutschen Volkes for financial support.
\end{acknowledgements}

%
%

\bibliography{refsphil.bib}   

%
%

\end{document}